# Migration-Driven Hierarchical Crystal Defect Aggregation — Symmetry and Scaling Analysis

Yuri G. Gordienko

G.V.Kurdyumov Institute for Metal Physics, National Academy of Sciences of Ukraine,

36 Academician Vernadsky Blvd, Kyiv-142, UA-03680, Ukraine

gord@imp.kiev.ua

**Abstract.** Recently the hierarchical defect substructures were observed experimentally in several metals and alloys before and after fracture. The general models of crystal defect aggregation with appearance hierarchical defect substructures are proposed and considered in the wide range of scales. Their general group analysis is performed, and symmetries of the governing equations are identified. The models of defect aggregate growth are considered for several partial cases and compared with classical Lifshitz-Slyozov-Wagner theory of coarsening, Leyvraz-Redner scaling theory of aggregate growth, etc. The reduced equations of new models are generated and solved, and the general scaling solutions are given. The results obtained are illustrated by preliminary simulations.

## Introduction

In materials science the fractured surfaces demonstrate the self-affine geometry on many scales (see, for example, from earlier works on metals after plastic deformation [1, 2] to recent results on high-strength low-alloy steels after bending-torsion fatigue [3], fracture roughness and toughness of concrete [4], discrete nature of crack propagation to the fractal geometry of the crack [5], etc.). Moreover, the fractal analysis of fractured surfaces by projective covering and box-counting method shows that the fractured surface can be depicted not only by one fractal dimension, but also by multifractal spectrum [6]. At the same time, surface roughness profiles of periodically deformed Al [7,8], slip line morphology in Cu [9-11], and dislocation patterns in Cu after tensile [12] also demonstrate the self-similar features on many scales. Several models and theories were proposed to describe and explain the scale-invariant behavior of crystal defect aggregations that possibly lead to self-affine geometry of fractured surfaces [13]. For example, the statistical model of noise-induced transition from homogeneous dislocation arrangements to scale-invariant structure with hyperbolic distribution was proposed recently [12].

The hierarchical defect substructures that were observed experimentally in deformed metals and alloys appear as a result of some aggregation processes among solitary crystal defects. Many aggregation phenomena in other physical processes take place by exchange of solitary agents (particles) between their aggregates (clusters): phase ordering [14], atom deposition [15], stellar evolution [16], growth and distribution of assets [17], and city population even [18]. In such aggregation processes particles can leave one cluster and attach to another. Usually these exchange processes are described by an exchange rate kernel $K(i, j)$, i.e. the rate of transfer of particles from a cluster of size $i$ (detaching event) to a cluster of size $j$ (attaching event). Generally, the rate of monomer particle exchange between two clusters depends on their active interface surfaces that are dependent on their sizes, morphology (line, plane, disk, sphere, fractal, etc), probability of detaching and attaching events, etc.

Sometimes there is the preferable direction for exchanges, i.e. with asymmetric exchange kernels, $K(i, j) \neq K(j, i)$, like in coalescence processes in Lifshitz-Slyozov-Wagner theory [14], where big clusters "eat" smaller ones. The exchange rate kernel $K(i, j)$ is defined by the product of the rate at which particle detach from a cluster of size $i$ and the rate at which this particle reach another cluster of size $j$. In Leyvraz-Redner scaling theory of aggregate growth [18] cities $A_i$ of size $i$ evolve according to the following rule:

$$A_i + A_j \xrightarrow{K(i,j)} A_{i-1} + A_{j+1}, \qquad (1)$$

where $K(i, j)$ is the exchange rate. That is, monomer particle (one person) leaves some of cities $A_i$ of population $i$ and arrive to some of cities $A_j$ of population $j$. This can be considered as generalized rule for the theory of growth and distribution of assets [17], if one can assume that $A_i$ are persons with asset volume of $i$.

Below the idealized general model of aggregate growth is proposed on the basis of this approach. The main aim of the work is to find the most profound features of aggregation kinetics, and factors that can influence the self-affine nature of the aggregating system of solitary agents (particles) and their aggregates (clusters). In this context, the numerous complex details of the real crystal defect aggregation processes will be hidden behind the idealized and simplified conditions only to emphasize the most general precursors of scale-invariant behavior of such complex systems.

**The general model**

Here detaching and attaching processes are considered *separately* that in the general case could be characterized by different rates. Consequently, the different *detach* product kernel $K_d(n) = k_d S(n)$ and *attach* product kernel $K_a(n) = k_a S_a(n)$ are taken into account, where $k_d$ and $k_a$ are the measures of activation of detaching and attaching processes, $n$ is the number of particles in a cluster. In natural processes $k_d$ is usually determined by energy barrier for detachment from cluster and $k_a$ — by probability for attachment of migrating particle to another cluster which in turn determined by kind of migration (instant hops from cluster to cluster, ballistic motion, random walking, or their combinations). $S_d(n) = s_d n^a$ and $S_a(n) = s_a n^b$ are the active surfaces of clusters, where $a$ and $b$ — exponents depending on the morphology of cluster (for example $a = 1$ for linear clusters and $a = 2/3$ for spherical clusters, and $a = b$ in the simplest case of clusters with the same morphology), $s_d$ and $s_a$ — the constants depending on the morphology of cluster and geometry of neighborhood (for example $s_d = 1$ for linear aggregates and $s_d = \sqrt[3]{36p}$ for spherical aggregates, and $s_d = s_a = s$ in the simplest case of clusters with the same morphology and neighborhood). The portion of clusters $f(n,t)$ with $n$ particles at time $t$ evolves according to the following equation:

$$\frac{\partial f(n,t)}{\partial t} = K_d(n+1)f(n+1,t) + K_a(n-1)f(n-1,t) - K_d(n)f(n,t) - K_a(n)f(n,t). \qquad (2)$$

The general model is based on the assumption that the migration time (movement of a particle from one cluster to other) is much lower than the detachment (or attachment) time of a particle from (to) a cluster. It should be taken into account that the more consistent system of evolution equations and more realistic coupling between migration and detachment (attachment) times should be used for rigorous comparison of this general model and the aforementioned natural processes.

**Transformation to the general Fokker-Planck equation.** In an asymptotic regime of high values of $n$ we can get

$$\frac{\partial f(n,t)}{\partial t} \approx \frac{\partial (D_1(n) f(n,t))}{\partial n} + \frac{\partial^2 (D_2(n) f(n,t))}{\partial n^2}. \qquad (3)$$

That is the one-variable Fokker-Planck equation in general form with time-independent drift $D_1(n) = K_d(n) - K_a(n) = n^a s(k_d - k_a)$ and diffusion $D_2(n) = K_a(n) = n^a s k_a$ coefficients [19]. The main difference between this formulation and other well-known models is that here the more general scenario with detaching and attaching as separate events is taken into account that allows us to express explicitly the availability of drift and diffusion terms. Moreover, it will allow to bring to light the common symmetry properties and propose the ways to find the exact non-stationary solutions (see below).

**Scaling ansatz for some partial cases.** It is relatively easy to show that for $k_d = k_a$ the one-variable Fokker-Planck equation (Eq. 3) is invariant under any stretching (one-parameter dilation group) transformations $e^a$ of independent and dependent variables: $\tilde{t} = (e^a)^{2-a} t$, $\tilde{n} = e^a n$, $\tilde{f} = f$ [20]. It means that one can use the scaling ansatz $f(n,t) = c(x)$, where $x = n t^{2-a}$, then derive the reduced set of combinations of the basic variables that are invariant under the group (similarity variables) and without loss of generality expect a solution in the form

$$f(n,t) \sim F\left(\frac{n}{t^{\frac{1}{2-a}}}\right). \qquad (4)$$

It allows us to reduce (Eq. 3) and analyze its properties in some partial cases which are of great interest in the context of their practical applications to description of the aforementioned natural processes [1,4,5].

*No drift: equiprobable exchange by monomers.* The case $k_d = k_a$ corresponds to the equiprobable activation of detaching and attaching processes and it means the absence of the drift term. For example, in coalescence processes in Lifshitz-Slyozov-Wagner theory [14] the equiprobable activation of detaching and attaching processes is provided by the equality between the rate of evaporation from cluster and the probability to reach another cluster. In Ispolatov-Krapivsky-Redner model of asset exchanges [17] it corresponds to the equal probability for some agent to give or get some assets. In Leyvraz-Redner scaling theory of aggregate growth of city population [18] it corresponds to equal probability for a person to departure from a city and to arrive to another city.

*Minimum active surface (singularity).* The case $a = 0$ corresponds to the clusters with the minimum active surface ("singularity"' of the active surface), namely for clusters with *constant* numbers of active particles independent of the whole number of particles $n$ in it. For example, ends of line cluster can be its 2 active points for detaching and attaching. Such configurations take place in queues (2 active points), stacks (1 active point), etc. In the context of plastic deformation of metals, "pile-up" aggregation of dislocations of regular crystalline structure can be depicted by this scenario (see Fig. 1) [20].

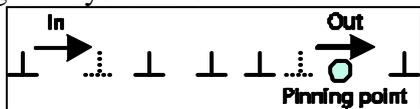

Fig. 1. Example of minimum active surface --- pile-up of dislocations: it is possible to go in/out pile-up *only* through left/right ends of the pile-up.

The condition $k_d = k_a$ corresponds to the equiprobable activation of detaching and attaching dislocations from opposite sides of pinning point and it means the absence of the drift term in (Eq. 3). Finally one can get the well-known heat equation

$$\frac{\partial f(n,t)}{\partial t} = sk_a \frac{\partial^2 f(n,t)}{\partial n^2} \quad \text{with solution} \quad f(n,t) \sim F\left(\frac{n}{\sqrt{t}}\right). \tag{5}$$

In the reality $k_d \neq k_a$, and the aggregation process is described by the homogeneous heat equation with space-independent drift and diffusion coefficients. It is well-known fact that it leads to "a diffusive-like kinetic universality class" [20].

*Maximum active surface.* The case $a = 1$ corresponds to the clusters with the maximum active surface, namely for clusters where *any* particle could be detached and free particle could attached to *any* place. In solid state physics, "wall" aggregation of dislocations of regular crystalline structure can be depicted by this scenario (see Fig. 2) [20]. In Leyvraz-Redner scaling theory of aggregate growth of city population [18] it corresponds to linear dependence of arrival or departure rates as a function of city population. Similarly, in Ispolatov-Krapivsky-Redner model of asset exchanges [17] it corresponds to linear dependence to give or get some assets as a function of the whole volume of assets for some agent.

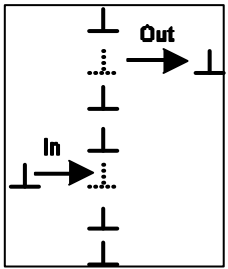

Fig. 2. Example of maximum active surface --- wall of dislocations: it is possible to go in/out wall in *any* place.

For periodic tensile deformation again one can assume that $k_d = k_a$, and it corresponds to the equiprobable activation of detaching and attaching dislocations from opposite sides of wall and it means the absence of the drift term in (Eq. 3). Finally one can get

$$\frac{\partial f(n,t)}{\partial t} = sk_a \frac{\partial^2 (n f(n,t))}{\partial n^2} \quad \text{with solution} \quad f(n,t) \sim F\left(\frac{n}{t}\right). \tag{6}$$

*Intermediate active surface.* The case $0 < a < 1$ corresponds to the clusters with the bulk volume of particles shielded by active surface. For example, circumference of disk, perimeter of fractal, outer layer of sphere can be the correspondent active surfaces for detaching and attaching. In solid state physics, such configurations take place in regular arrangements (compact like voids and spare like fractals) of point-like defects of crystalline structure [20].

**Symmetry analysis**

**Extended symmetries of the general Fokker-Planck equation.** The complete classification of the extended symmetries of this Fokker-Planck equation was proposed recently with three cases [21]:

(I) the symmetry group with 4 non-trivial symmetry generators (that is locally isomorphic to the group of the heat equation);

(II) the symmetry subgroup of (I) with 2 non-trivial symmetry generators;
(III) the trivial symmetry only.

Thus, (Eq. 3) could be reduced to one with drift $\tilde{D}_1(n) = D_1(n)/D_2(n) = (k_d - k_a)/k_a$ and diffusion $\tilde{D}_2(n) = 1$ coefficients not changing the symmetry properties of the original equation [22]. Because the necessary and sufficient conditions are satisfied by these coefficients for case I [21], the migration-driven aggregate growth for all possible morphology types of clusters can be characterized by transformed one-variable Fokker-Planck equation which symmetry is equivalent to symmetry of the heat equation.

**On scaling and exact non-stationary solutions.** The transformation of the rate equation (Eq. 2) to the general Fokker-Planck equation (Eq. 3) is very useful and convenient for practical purposes. The stationary solutions of the general Fokker-Planck equation (Eq. 3) could be found relatively easy, but usually it is hard task to find the exact non-stationary solution for the one-dimensional Fokker-Planck equation with arbitrary drift and diffusion coefficients. However, recent results on the relations between the one-dimensional Fokker-Planck equation and the Schrödinger equation allows us to obtain the transformed equation, which can be easily treated to find the exact non-stationary solutions.

*Transformation to the normalized form of Fokker-Planck equation.* By substitution $x = \int D_2^{-1/2}(n)\,dn$ to (Eq. 3) the following reduced equation could be obtained

$$\frac{\partial f(x,t)}{\partial t} = \frac{\partial^2 f(x,t)}{\partial x^2} + A(x)\frac{\partial f(x,t)}{\partial x} + B(x)f(x,t),$$

where

$$A(x) = \frac{D_1(x)}{\sqrt{D_2(x)}} + \frac{2}{D_2(x)}\frac{\partial D_2(x)}{\partial x}; \quad B(x) = \frac{1}{\sqrt{D_2(x)}}\frac{\partial D_2(x)}{\partial x} + \frac{1}{D_2(x)}\frac{\partial^2 D_2(x)}{\partial x^2}. \tag{7}$$

And by substitution $u = f\exp\{\tfrac{1}{2}\int A(x)dx\}$ to (Eq. 5) the normalized form of Fokker-Planck equation could be obtained:

$$\frac{\partial u(x,t)}{\partial t} = \frac{\partial^2 u(x,t)}{\partial x^2} + I(x)u(x,t), \text{ where } I(x) = B(x) - \frac{A^2(x)}{4} - \frac{1}{2}\frac{\partial A(x)}{\partial x}. \tag{8}$$

*On possibility to calculate exact non-stationary solutions.* The normalized form of Fokker-Planck equation (Eq. 8) is actually Schrödinger equation with potential $V(x)$, which for the case (I) [21] should have quadratic potential. It can be easily analyzed by the standard procedures [23], for example the Green function can be calculated for all aforementioned theories, from which the non-stationary solutions with arbitrary initial condition can be obtained, if the initial and boundary conditions will be strictly formulated [20].

In the view of the possible transformations (Eq. 4) and (Eq. 7) (that are dependent on morphology of clusters, active surface, and geometry of neighborhood) the different set of solutions are possible. That is why their scaling (if it is available) in the aforementioned theories (in an asymptotic regime of high values of $n$) is highly dependent on geometry and morphology of clusters, and kind of migration (ballistic, diffusive, arbitrary) which are very important for estimation of the effective active surface [20].

The amazing fact is the afomentioned transformations lead to Schrödinger equation (Eq. 8) that has the same extended group (up to local isomorphism) as "free Schrödinger equation" with $V(x) = 0$ [24]. It sufficiently simplifies the scaling analysis and calculation of exact non-stationary solutions for aforementioned theories.

**Illustrative simulations of several primitive models**

The two aforementioned primitive models of cluster aggregation were simulated by Monte Carlo method to illustrate the different cluster distributions in different aggregation kinetics. Several initial configurations of $10^5$ particles were used in simulations: (A) all solitary particles, i.e. $10^5$ "clusters" with 1 particle in each of them; (B) $10^4$ clusters with 10 particles in each of them; (C) $10^3$ clusters with $10^2$ particles in each of them.

*Model 1 — minimum active surface (singularity).* It corresponds to "pile-up" aggregation of dislocations in crystals (see Fig. 1) with $k_d \neq k_a$. The kinetics of rearrangements from three initial configurations is shown in Fig. 3. After some transition regime the broad peak appears, which corresponds to the average cluster size <N> (i.e. the mean number of particles in each pile-up) appears. Despite the different initial configurations after some time sweeps the initial "singular" (like Dirac delta function) distributions evolve to similar distributions with broad peaks (Fig. 3).

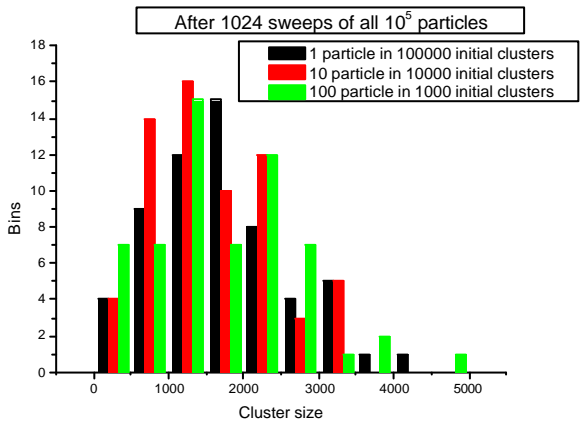

Fig. 3. The distributions of cluster size (the number of particles) for different initial *pile-up* configurations.

*Model 2 — maximum active surface.* It corresponds to "wall" aggregation of dislocations in crystals (see Fig. 2). The kinetics of rearrangements from initial configurations is shown in Fig. 4. Preliminary simulations show that after some transition regime the average cluster size <N> changes *nearly* linearly (<N> ~ $t^B$, where $B = 0.95 \pm 0.1$) as a function of time steps, that corresponds to the scaling behavior predicted by (Eq. 5).

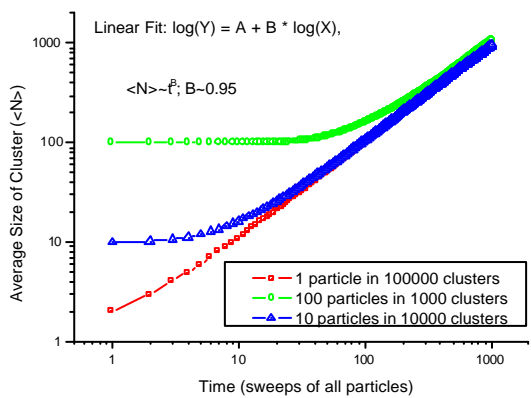

Fig. 4. The average cluster size <N> (the number of particles) as a function of time steps for different initial *wall* configurations: squares — all solitary particles (configuration A), triangles — many small clusters (configuration B), circles — several big clusters (configuration C).

Despite the deceptive term "average cluster size", the initial "singular" configurations of walls after some time sweeps evolve to scale-free distributions (Fig. 5), that is different from pile-up distributions with broad peaks for the same time. That is why "average cluster size" notion seems to be meaningless for this kind of distributions without distinctive peaks.

That is why in practice to estimate the level of hierarchy of defect substructures and related scale of damage it is necessary to measure not only the integral "averaged" characteristics of defect ensembles (spatial density, average size, etc.), but also to define more specific characteristics, related with their size distribution, spatial distribution, time evolution, etc. For example, scaling analysis of experimental defect aggregates distributions will allow to determine their intrinsic symmetry and bring to light the corresponding aggregation scenario.

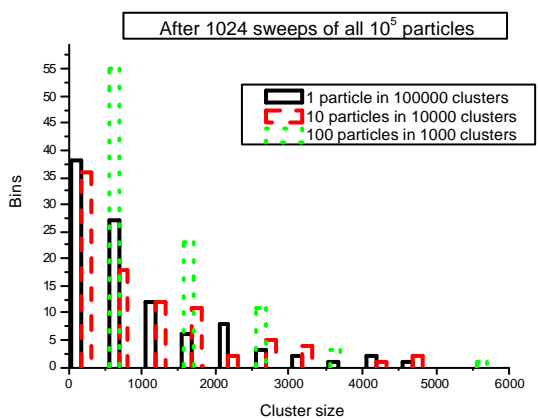

Fig. 5. The distributions of cluster size (the number of particles) for different initial *wall* configurations

It should be noted that to the moment some details of the simulation process cannot be compared directly with the theoretical results for large $n$ and $t$ because simulations were carried out for relatively low number of aggregating particles ($<10^6$) and in the limited time range ($<10^4$ steps). That is why the bigger simulations are under way now to check the exact solutions and make more reliable conclusions [20].

**Summary**

The idealized general approach to characterize defect aggregate growth was proposed that allowed to find the scaling characteristics of some aggregation scenarios. Some factors that can influence the self-affine nature of the aggregating system of solitary agents (particles) and their aggregates (clusters) were brought to light: the rate of monomer particle exchange between two clusters, cluster geometry, cluster morphology, and kind of migration (ballistic, diffusive, etc.) of monomer particles. It was shown that the simplified aggregate growth can be described by the one-variable Fokker-Planck equation in general form with time-independent drift and diffusion coefficients. In some aggregation scenarios it is invariant under any stretching transformations of independent and dependent variables (one-parameter dilation group). It could be transformed to the equivalent heat equation without changing the symmetry properties of the original equation, and then it could be transformed to the Schrödinger equation. This means that Green functions can be easily calculated, from which the non-stationary solutions with arbitrary strict initial and boundary conditions can be obtained. From the practical point of view the scaling analysis of experimental distributions of defect aggregates will allow to determine their intrinsic symmetry and bring to light the corresponding aggregation scenario.

**Acknowledgements**

The work was partially supported by INTAS grant 04-80-7078 and EDGeS (Enabling Desktop Grids for e-Science) Project RI211727 of 7th FP.

**References**

[1] B.B. Mandelbrot: Nature, 308 (1984), p.721.

[2] K. Banerji: Metall Trans A, 19 (1988), p. 961.

[3] K. Slámecka, P. Poníźil, J. Pokluda: Mater. Sci. and Eng.: A, 462, Issues 1-2 (2007), p. 359.

[4] M.A. Issa, M.A. Issa, Md.S. Islam, A. Chudnovsky: Eng. Fract. Mech., 70 (2003), p.125.

[5] M.P. Wnuk, A. Yavari: Eng. Fract. Mech., 75 (2008), p. 1127.

[6] S. Stach, J. Cybo: Mater. Charact., 51 (2003), p. 79; *ibid*. p. 87.

[7] Yu.G. Gordienko, E. Zasimchuk: Proceedings of the SPIE, 2361 (1994), p. 312.

[8] E.E. Zasimchuk, Yu.G. Gordienko, R.G. Gontareva, I.K. Zasimchuk: J. Mater. Eng. and Perf., 12 (2003), p. 68; Yu.G. Gordienko, et al: Adv. Eng. Mater., 8, Issue 10 (2006), p. 957.

[9] T. Kleiser, M. Bocek, Z. Metalkde 77 (1986) 587.

[10] H. Neuhäuser, in: L.P. Kubin, G. Martin (Eds.), *Nonlinear Phenomenon in Materials Science*, vol. 3 & 4 of Solid State Pheneomenon (Trans. Tech. Publications, Switzerland 1988), p. 407.

[11] M. Zaiser, F.M. Grasset, V. Koutsos, E.C. Aifantis, Phys. Rev. Lett. 93 (2004) 195507.

[12] P. Hähner, K. Bay, M. Zaiser, Phys. Rev. Lett. 81 (1998) 2470.

[13] G. Ananthakrishna: Physics Reports, 440 (2007), p. 113.

[14] I.M. Lifshitz and V.V. Slyozov: Zh. Eksp. Teor. Fiz., 35 (1959), p. 479 [Sov. Phys. JETP, 8, (1959), p. 331]; J. Phys. Chem. Solids, 19 (1961), p. 35.

[15] A. Zangwill: *Physics at Surfaces* (Cambridge University Press, New York 1988).

[16] S. Chandrasekhar: Rev. Mod. Phys., 15 (1943), p. 1.

[17] S. Ispolatov, P.L. Krapivsky, and S. Redner: Eur. Phys. J., B 2 (1998), p. 267.

[18] F. Leyvraz and S. Redner: Phys. Rev. Lett., 88 (2002) 068301.

[19] A.D. Fokker: Ann.Phys., 43 (1914), p. 810; M.Planck: Preuss. Akad. Wiss. Phys. Math. K1 (1917), p. 325.

[20] Yu. Gordienko (to be published).

[21] G. Cicogna and D. Vitali: J.Phys.A: Math.Gen., 23 (1990), p. L85.

[22] E.B. Dynkin: *Markov Processes* (Moscow, Fizmatgiz 1963).

[23] T. Miyazawa: Phys. Rev. A, 39 (1989), p. 1447.

[24] U. Niederer, Helv. Phys. Acta, 46 (1973), p. 191; C.P. Boyer, Helv. Phys. Acta, 47 (1974), p. 589.